\documentclass{article}
\usepackage{amsfonts}
\usepackage{ijcai24}

\usepackage{times}
\usepackage{soul}
\usepackage{url}
\usepackage[hidelinks]{hyperref}
\usepackage[utf8]{inputenc}
\usepackage[small]{caption}
\usepackage{graphicx}
\usepackage{amsmath}
\usepackage{amsthm}
\usepackage{booktabs}
\usepackage{algorithm}
\usepackage{algorithmic}
\usepackage[switch]{lineno}
\usepackage{xcolor}

\usepackage{bm}

\usepackage{graphicx}
\usepackage{float}
\usepackage{subfigure}
\usepackage{amssymb}

\title{Advancing Medical Image Segmentation via Self-supervised Instance-adaptive Prototype Learning}

\author{
Guoyan Liang$^{1}$
\and
Qin Zhou$^{2}$ 
\and
Jingyuan Chen$^1$\and
Zhe Wang$^{2}$\thanks{Corresponding Authors.} \And
Chang Yao$^{1 *}$\\
\affiliations
$^1$Zhejiang University, Hangzhou, China\\
$^2$East China University of Science and Technology, Shanghai, China\\
\emails
\{guoyanl, jingyuanchen, changy\}@zju.edu.cn,
\{sunniezq, wangzhe\}@ecust.edu.cn
}

\begin{document}

\maketitle

\begin{abstract}
    Medical Image Segmentation (MIS) plays a crucial role in medical therapy planning and robot navigation. Prototype learning methods in MIS focus on generating segmentation masks through pixel-to-prototype comparison. However, current approaches often overlook sample diversity by using a fixed prototype per semantic class and neglect intra-class variation within each input. In this paper, we propose to generate instance-adaptive prototypes for MIS, which integrates a common prototype proposal (CPP) capturing common visual patterns and an instance-specific prototype proposal (IPP) tailored to each input. To further account for the intra-class variation, we propose to guide the IPP generation by re-weighting the intermediate feature map according to their confidence scores. These confidence scores are hierarchically generated using a transformer decoder. 
   Additionally we introduce a novel self-supervised filtering strategy to prioritize the foreground pixels during the training of the transformer decoder. Extensive experiments demonstrate favorable performance of our method.
\end{abstract}

\section{Introduction}
Medical image segmentation (MIS) aims to divide an image into meaningful parts, representing different semantic classes. It serves as a fundamental requirement for various tasks such as cancer diagnosis, radiation therapy dosage control and surgical robot navigation.
Recognizing the significant resemblance between MIS and the task of clustering pixels into semantic segments, recent researchers have focused on prototype-based methods. By learning representative prototype features, which serve as cluster centers for each semantic class, the MIS objective is reformulated as pixel-to-prototype matching. Existing prototype-based MIS methods primarily concentrate on learning representative prototypes, which remain fixed upon completion of the training process \cite{zhou2022rethinking}. In order to capture the intra-class variation within each semantic class, prior studies have explored the use of local prototypes \cite{yu2021location,wang2022few}. However, these prototypes are predominantly generated based on spatial grids or pre-segmentation techniques such as superpixels. 
Recently, the introduction of mask transformers \cite{Cheng_2022_CVPR,yu2022k,yuan2023devil} has allowed for the direct learning of object queries using a transformer decoder. By considering these object queries as cluster centers, we can obtain the segmentation mask by assigning pixels to clusters and mapping clusters to their corresponding semantic classes. However, as mentioned in \cite{Cheng_2022_CVPR,yu2022k,yuan2023devil}, without proper guidance, existing mask transformers may introduce a bias towards focusing on the background.

To tackle the aforementioned challenges in prototype-based segmentation, we introduce a novel framework called Self-supervised Instance-adaptive Prototype Learning (referred as SIPL) for MIS. Specifically, we propose an Instance-adaptive Prototype Learning (referred as IPL) scheme that adaptively learns prototypes tailored to different inputs. In IPL, the final class-specific prototypes are generated using a combination of common prototype proposals (CPPs), which capture the common visual patterns of each class, and instance-specific prototype proposals (IPPs), which dynamically emphasize the instance-specific visual characteristics of the current input.

To further account for the intra-class variation during prototype generation, we propose a novel hierarchical Self-supervised Mask Generation (referred as SMG) module to hierarchically generate pseudo masks with confidence scores to guide the generation of IPPs. These confidence scores are used to re-weight the intermediate feature map, indicating the pixel-wise importance in contributing to the IPPs for each class. In SMG, a transformer decoder is employed to generate query embeddings from the multi-scale features obtained from the pixel decoder.
To mitigate background bias in the SMG module, we incorporate a self-supervised filtering strategy that prioritizes foreground pixels during the learning of query embeddings. To improve the accuracy of the confidence scores, We further introduce an auxiliary loss to supervise the training process of the transformer decoder. 

The overall contributions of our method can be summarized as follows.
\begin{itemize}
    \item We propose a novel framework to learn instance-adaptive prototypes for MIS, which incorporates common and instance-specific prototype proposals to capture the common and instance-specific visual patterns for each class.
    \item We propose a self-supervised mask generation module to guide instance-specific prototype proposal generation by re-weighting intermediate features, where we incorporate a novel self-supervised filtering strategy to prioritize foreground pixels during pseudo mask generation.
    
    \item Extensive experiments across three challenging MIS tasks demonstrate favorable performance of our method.
    
\end{itemize}

\section{Related Works}
\subsection{Medical Image Segmentation}
Early methods in medical image segmentation (MIS) primarily rely on contour-based and traditional machine learning algorithms \cite{tsai2003shape}. 
In contrast, contemporary MIS approaches predominantly leverage deep CNNs and transformer models.
U-Net initially introduced for MIS based on CNN, stands out for its simplicity and superior performance \cite{ronneberger2015u}.
Subsequently, various U-Net variants have emerged for 2D MIS such as Res-UNet \cite{xiao2018weighted}, U-Net++ \cite{zhou2018unet++}, and UNet3+ \cite{huang2020unet}.
Transformer-based methods have emerged due to the impressive speed-accuracy performance of the Vision Transformer (ViT) in various visual tasks \cite{dosovitskiy2020image}.
Swin Transformer has notably excelled by introducing an efficient and effective hierarchical vision transformer that leverages the shifted windows mechanism \cite{liu2021swin}. Building upon this paradigm, Swin-unet incorporates a U-shaped Encoder-Decoder architecture with skip connections, further advancing the field \cite{cao2022swin}.
Segtran proposes a novel positional encoding scheme for transformers, which imposes a continuity inductive bias for images and achieves high segmentation accuracy with good cross-domain generalization capabilities \cite{li2021medical}.
Some recent works explore hybrid approaches that demonstrate improved segmentation performance compared to pure CNNs and transformers-based counterparts. UNETR seamlessly integrates the long-range spatial dependencies of transformers with the inductive bias of CNNs within a ``U-shaped" encoder-decoder architecture \cite{hatamizadeh2022unetr}.
nnFormer strategically combines CNN and transformer-based blocks, interleaving them to harness the complementary strengths of each \cite{zhou2021nnformer}.
UNETR++ introduces an innovative Efficient Paired Attention (EPA) block, skillfully capturing enriched inter-dependent spatial and channel features by employing both spatial and channel attention in two distinct branches \cite{abdelrahman2022unetr++}.

\subsection{Prototype Based Methods}

Prototype learning assigns query embeddings to class centers in the network's output space, typically defined as the mean vector of the training set for each class \cite{mensink2013distance}. 
Prototype-based networks exhibit a straightforward output layout \cite{wen2016discriminative} and demonstrate fast generalization to new classes. As a result, they often concentrate on the few-shot regime, \cite{boney2017semi,snell2017prototypical} 
or update the prototypes online based on mini-batches \cite{guerriero2018deepncm}.
Matching Network is proposed to leverage learned prototypes for one-shot learning, capturing class features during training and demonstrating robust classification during testing \cite{vinyals2016matching}. 
Prototypical Network utilizes prototype spaces to compute distances between input samples and class prototypes for few-shot learning, proving effective in scenarios with limited labeled data. 
\cite{santoro2016meta} incorporates memory augmentation to store task-specific prototype information, facilitating rapid adaptation to new tasks. 
Relation Networks \cite{sung2018learning} emphasizes modeling relationships between input samples and class prototypes for enhanced few-shot learning performance.
Additionally, \cite{chen2019closer} refines prototype-based methods and explore improved strategies to leverage limited training data for more accurate classification. While existing prototype-based methods have made significant progress, they often lack the flexibility to dynamically adapt to different inputs \cite{zhou2022rethinking}. In contrast, our method focuses on learning instance-adaptive prototypes, resulting in improved performance in MIS. 

\section{Method}
In this section, we will provide an overview of our proposed method. We start by discussing the background and limitations of existing prototype-based segmentation methods in Sec.~3.1. Then, we present the Self-supervised Instance-adaptive
Prototype Learning (SIPL) framework in Sec. 3.2. We further explain the Instance-adaptive Prototype Learning (IPL) strategy in Sec. 3.3 and the hierarchical Self-supervised Mask Generation (SMG) module in Sec. 3.4.


\subsection{Revisiting Prototype Based Segmentation}

\begin{figure*}[t]
\centering
\includegraphics[width=0.8\textwidth]{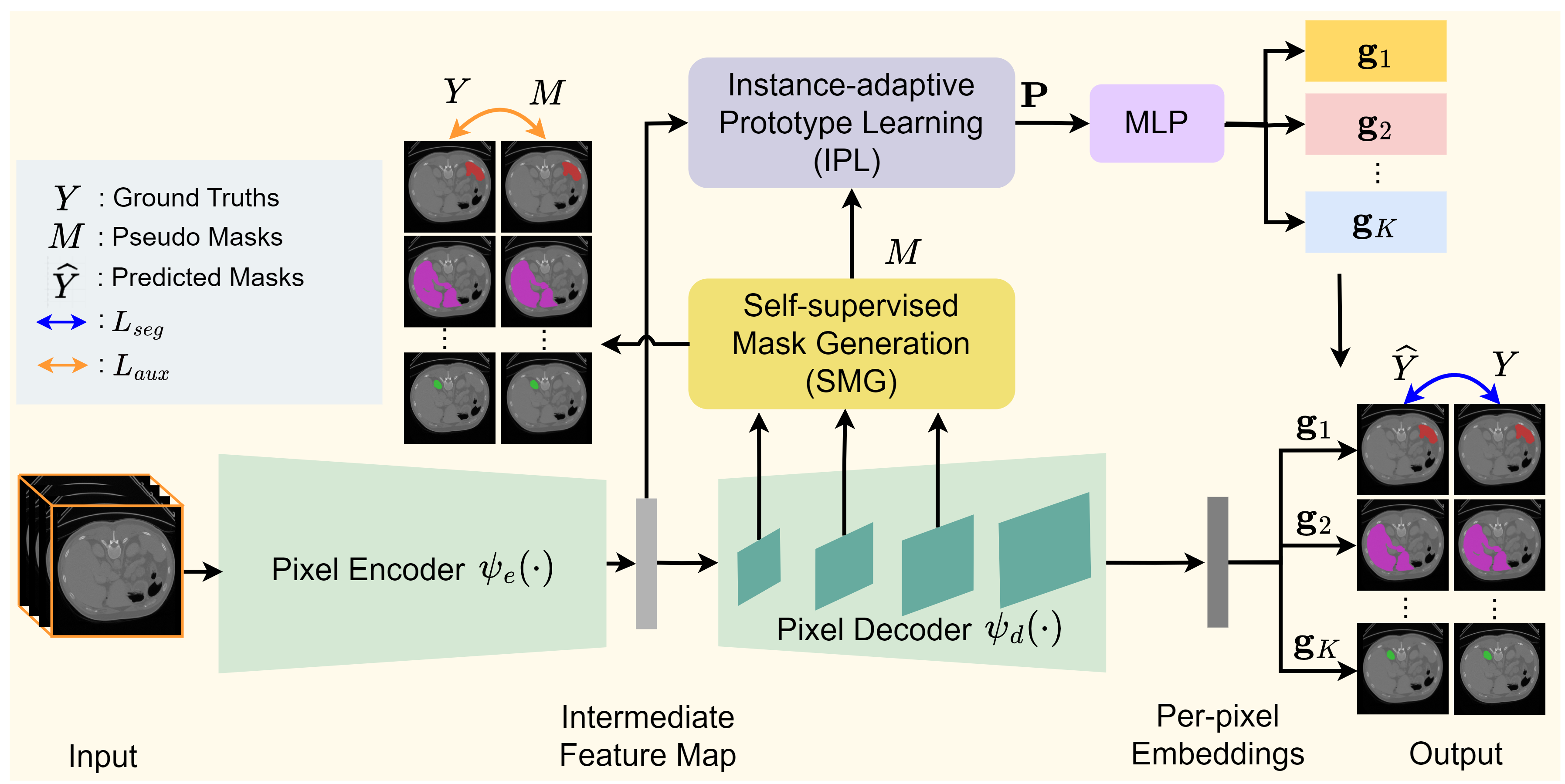}
\caption{Overview of our proposed SIPL framework. }
\label{fig1:env}
\end{figure*}

Prototype learning methods in Medical Image Segmentation (MIS) focus on generating segmentation masks through pixel-to-prototype comparison. These approaches typically learn one prototype feature for each class. Denote the learned prototypes as $\mathbf{G} = \{\mathbf{g}_k \in \mathbb{R}^d\}, k \in \{1,\cdots,K\}$, the $i$-th pixel embedding as $\mathbf{f}_i\in \mathbb{R}^d$, where $K$ is the number of classes and $d$ is the feature dimension. Then the probability that the $i$-th pixel belongs to the $k$-th class can be calculated as follows:
\begin{equation}
    p(k|\mathbf{f}_i) = \frac{exp(-<\mathbf{f}_i, \mathbf{g}_k>)}{\sum_{k'=1}^K exp(-<\mathbf{f}_i, \mathbf{g}_{k'}>)},
\end{equation}
where $<,>$ denotes the cosine similarity between two items. The existing prototype-based segmentation methods encounter challenges in two aspects: \textit{1) Fixed prototypes during inference that lack dynamic adaptation to different inputs. 2) Prototypes generated by averaging feature embeddings of each class, overlooking intra-class variation.} To overcome these limitations, we propose a novel framework that generates instance-adaptive prototypes for each input. In addition, we address intra-class variation by hierarchically generating pseudo masks with confidence scores, which serve as the pixel-wise importance in the instance-specific prototype proposal generation for each class. The subsequent section provides a detailed description of the network architecture.



\subsection{Overall Architecture}
Figure~\ref{fig1:env} illustrates the overall architecture of our method. Given the input 3D volume-mask pair $(\mathbf{V},Y)$, where $\mathbf{V} \in \mathbb{R}^{H \times W \times Z}$ is the 3D image patch, and $Y \in \mathbb{R}^{H\times W\times Z}$ refers to the corresponding segmentation mask. We utilize a 3D backbone comprising an encoder branch $\psi_e(\cdot)$ and a pixel decoder branch $\psi_d(\cdot)$ for segmentation mask prediction. Concretely, the input $\mathbf{V}$ is first fed into the encoder to obtain the intermediate feature map $\mathbf{F} \in \mathbb{R}^{\frac{H}{S} \times \frac{W}{S} \times \frac{Z}{S} \times d_i}$, where $S$ is the stride of the feature map, $d_i$ is the feature dimension of the intermediate feature map. Then $\mathbf{F} $ is fed into the pixel decoder branch $\psi_d(\cdot)$ to generate the output per-pixel embeddings $\mathbf{F}_o \in \mathbb{R}^{H \times W \times Z \times d}$. The final segmentation mask is predicated as the pixel-to-prototype comparison between $\mathbf{F}_o$ and the learned prototypes $\mathbf{G}$. 

To facilitate dynamic adaptation of learned prototypes to current input, we introduce the Instance-adaptive Prototype Learning (IPL) strategy, which combines Common Prototype Proposals (CPPs) and Instance-specific Prototype Proposals (IPPs) generation. The CPPs are learnable parameters obtained from the entire dataset, while IPPs are generated by reweighting the intermediate feature map  $\mathbf{F}$ using the pixel-wise confidence scores. The final prototype proposals $\mathbf{P}=\{\mathbf{p}_k\}, k \in \{1,\cdots,K\}$ are obtained by concatenating the CPPs and IPPs along the feature channel dimension. These proposals are then processed through a few fully connected layers (MLP) to obtain the final prototypes $\mathbf{G}$.


In order to account for the intra-class variation during the generation of IPPs, we introduce the Self-supervised Mask Generation (SMG) module to generate pixel-wise confidence scores for each class. This module utilizes a mask transformer to generate pseudo masks with confidence scores, where the pseudo masks are hierarchically aggregated from pixel-to-cluster and cluster-to-class assignments. 
To provide the IPL module with more accurate pseudo masks, we introduce the auxiliary loss ($L_{aux}$) between pseudo masks $M$ and ground truth masks $Y$. 

\subsection{Instance-adaptive Prototype Learning}

\begin{figure}[t]
\centering
\includegraphics[width=0.48\textwidth]{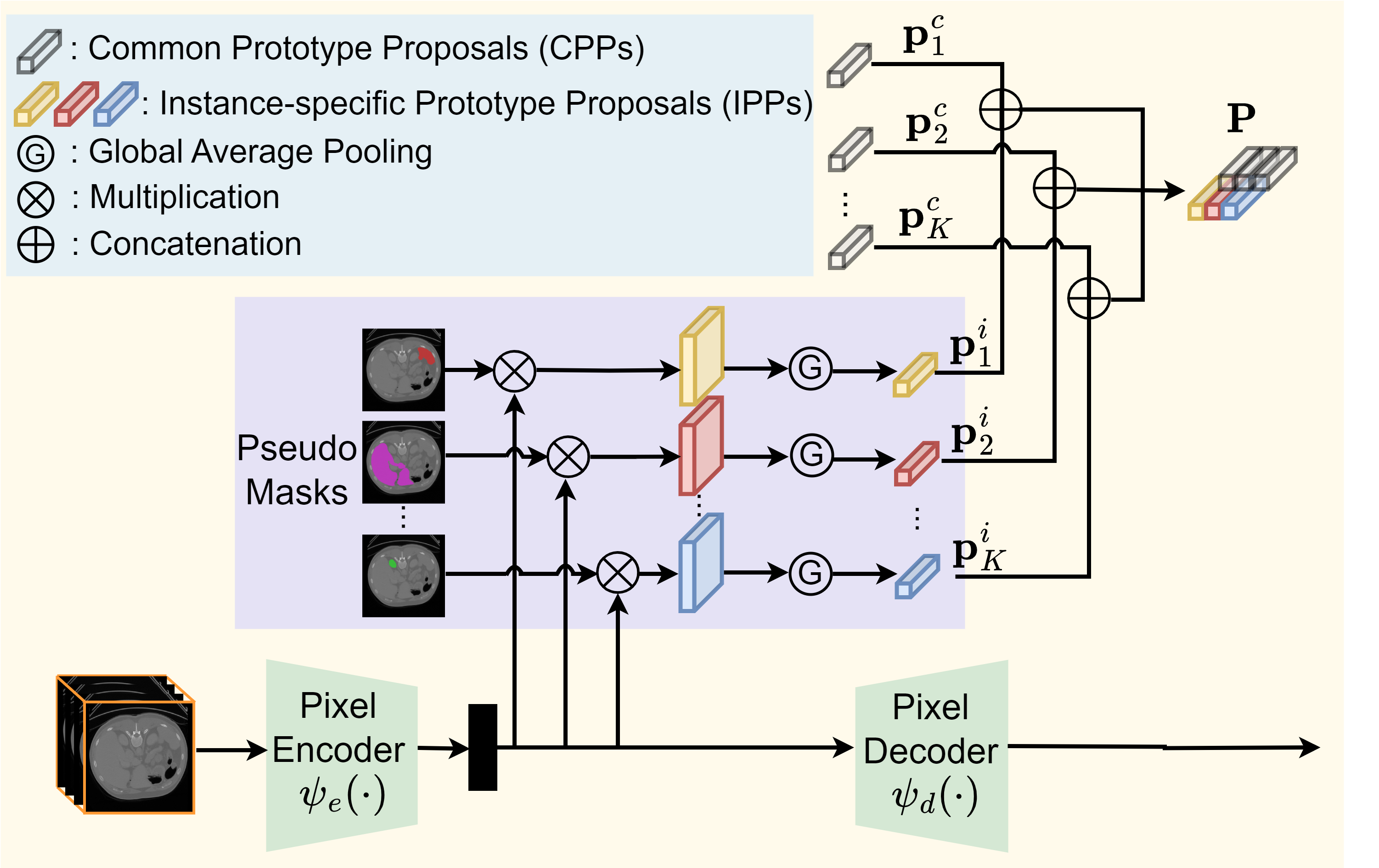}
\caption{Illustration of our instance-adaptive prototype learning (IPL) module. The IPL module consists of two components: common prototype proposal generation and instance-specific prototype proposal generation. The pseudo masks with confidence scores are hierarchically generated in a self-supervised manner to account for the intra-class variation, as described in Sec~\ref{sec:3.4}.}
\label{fig2:env}
\end{figure}

The overall workflow of IPL is illustrated in Figure \ref{fig2:env}. Formally, given the intermediate feature map $\mathbf{F} \in \mathbb{R}^{\frac{H}{S} \times \frac{W}{S} \times \frac{Z}{S} \times d_i}$ and pseudo masks from the final transformer decoder layer $M=\{M_k \in \mathbb{R}^{H \times W \times Z \times 1}\}$, $k\in \{1,2,\cdots, K+1\}$, where $M_k \in [0,1]$ denotes the predicted confidence scores of the $k$-th class. Then the instance-specific prototype proposal (IPP) for the $k$-th class $\mathbf{p}_k^i \in \mathbb{R}^{1 \times d_i}$ is generated as,
\begin{equation}
\mathbf{p}_k^i = GAP(M_k \otimes \mathbf{F}),
\label{eq:ipp}
\end{equation}
where $\otimes$ is the pixel-wise multiplication, $GAP()$ denotes global average pooling. In our method, pseudo masks $M$ are generated hierarchically from the self-supervised mask generation module (please refer to Eq.~(\ref{eq:m_l}) for details). Please note that the pseudo mask $M_k$ are reshaped to match the size of $\mathbf{F}$ in Eq.~(\ref{eq:ipp}).


In addition to the IPPs, our method includes a learnable common prototype proposal (CPP) for each class. This shared proposal captures the common visual dynamics of each class present in the entire training set. Concretely, denote the CPPs as $\mathbf{P}^c = \{\mathbf{p}_k^c\}$, $k \in \{1,2,\cdots,K\}$, where $\mathbf{p}_k^c \in \mathbb{R}^{1 \times d_i}$ denotes the prototype proposal of the $k$-th class. These CPPs are optimized by the overall learning objective and are randomly initialized in our approach. The IPPs and CPPs are then fused together to generate the final prototype $\mathbf{G}$. Formally, the final prototype belonging to the $k$-th class is obtained as,
\begin{equation}
\mathbf{g}_k = MLP(\mathbf{p}_k^i \oplus \mathbf{p}_k^c),
\end{equation}
where $\oplus$ denotes the concatenation operation, $MLP()$ refers to a few fully connected layers. Given the output per-pixel embeddings $\mathbf{F}_o \in \mathbb{R}^{H \times W \times Z \times d}$ from the pixel decoder branch and the learned prototype $\mathbf{g}_k \in \mathbb{R}^{d \times 1}$, the final segmentation prediction for the $k$-th class is formulated as,
\begin{equation}
 \widehat{Y_k} = \sigma(\mathbf{F}_o \circledast \mathbf{g}_k),
\end{equation}
where $\circledast$ denotes the matrix multiplication, $\sigma(\cdot)$ denotes the Sigmoid operation.
We adopt soft dice loss and binary cross-entropy to calculate the segmentation loss, which is mathematically formulated as,
\begin{equation}
    L_{seg} = 1 - \sum\limits_{k=1}^K (\sum_i \frac{  Y_{k,i} \widehat{Y_{k,i}} }{ Y_{k,i}^2 + \widehat{Y_{k,i}} ^2}  + \sum_i Y_{k,i} log \widehat{Y_{k,i}}  ).
\end{equation}
Here, $Y_{k,i}, \widehat{Y_{k,i}}$ correspond to the $i$-th pixel of the segmentation map for class $k$ in the ground truth and predictions, respectively.


\subsection{Self-supervised Mask Generation}
\label{sec:3.4}

\begin{figure}[h]
\centering
\includegraphics[width=0.48\textwidth]{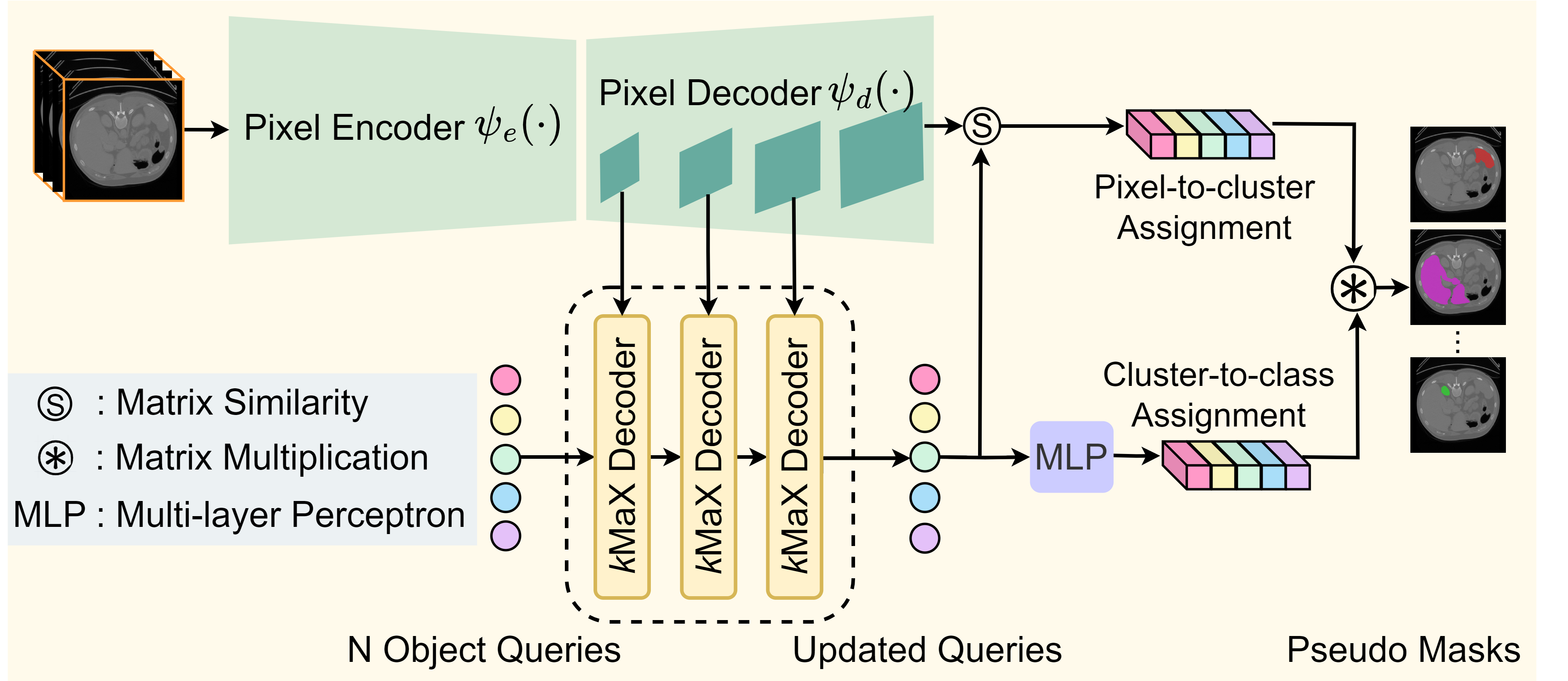}
\caption{Illustration of the self-supervised mask generation (SMG) module. In SMG, object queries are learned from a transformer decoder using multi-scale features from the pixel decoder. Pseudo masks are generated by hierarchically assigning pixels to clusters and clusters to classes.}
\label{fig3:env}
\end{figure}

Starting from DETR \cite{carion2020end} and MaX-DeepLab \cite{wang2021max}, the object queries, updated by multiple transformer decoders, are employed as mask embedding vectors. In our approach, we employ learnable object queries to capture intra-class variation, as shown in Figure \ref{fig3:env}. These object queries, which can be seen as clusters, are mapped onto the semantic class to generate pseudo masks with confidence scores. This process facilitates instance-adaptive prototype learning. Mathematically, denote the learnable object queries as $\mathbf{Q} \in \mathbb{R}^{N \times d_q}$, where $N$ ($N>K$) refers to the number of object queries, $d_q$ denotes the feature dimension of the object queries. Then $\mathbf{Q}$ is updated by kMax decoders \cite{yu2022k} using the multi-scale features from the pixel decoder. Denote the reshaped feature map from the $l$-th layer of the pixel decoder as $\mathbf{F}_{d}^l \in \mathbb{R}^{H_lW_lZ_l \times d_q}$, the randomly initialized query embeddings as $\mathbf{Q}^0$, then the updated query embeddings $\mathbf{Q}^l$ is calculated as, 
\begin{equation}
\begin{aligned}
    & \mathbf{A} = \mathop{argmax}\limits_{N}( \mathbf{Q}^{l-1}   \circledast (\mathbf{F}_d^l)^T), \\
     &\mathbf{Q}^l = MHSA(\mathbf{Q}^{l-1}) + \mathbf{A} \circledast \mathbf{F}_d^l ,
\end{aligned}
\end{equation}

where $\mathbf{Q}^{l-1}$ denotes the query embeddings obtained from the $l-1$ layer, $MHSA()$ denotes the multi-head self-attention. Based on the updated query embeddings $\mathbf{Q}^l$ from layer $l$, we can obtain the corresponding pixel-to-cluster assignment map $M^l_{pc}$ as,
\begin{equation}
    M^l_{pc} = sim(\mathbf{F}_{d}^l, \mathbf{Q}^l),
\end{equation}
where $sim(,)$ calculates the feature similarity between each pixel of $\mathbf{F}_{d}^l $ and each query embedding of $\mathbf{Q}^l$. The generated pixel-to-cluster assignment map $M^l_{pc}$ is of size $H_l\times W_l \times Z_l \times N$. Since the IPL module requires class-specific pseudo masks, we further introduce a cluster-to-class projection head to generate cluster-to-class projection matrix $M^l_{cc} \in \mathbb{R}^{N \times (K+1)}$,
\begin{equation}
    M^l_{cc} = MLP(\mathbf{Q}^l).
\end{equation}

We then aggregate the pixel-to-cluster assignments $ M^l_{pc}$ and the cluster-to-class mapping matrix $M^l_{cc}$ to obtain the pseudo masks for the $l$-th layer as,
\begin{equation}
    M^l = M^l_{pc} \circledast M^l_{cc}.
    \label{eq:m_l}
\end{equation}

$M^l$ undergoes softmax processing to get the normalized confidence scores. It should be noted that during the generation of $M^l $, the background class (i.e., $k = K+1$) is considered to obtain reasonable confidence scores using softmax.

\begin{figure*}[h]
\centering
\includegraphics[width=0.8\textwidth]{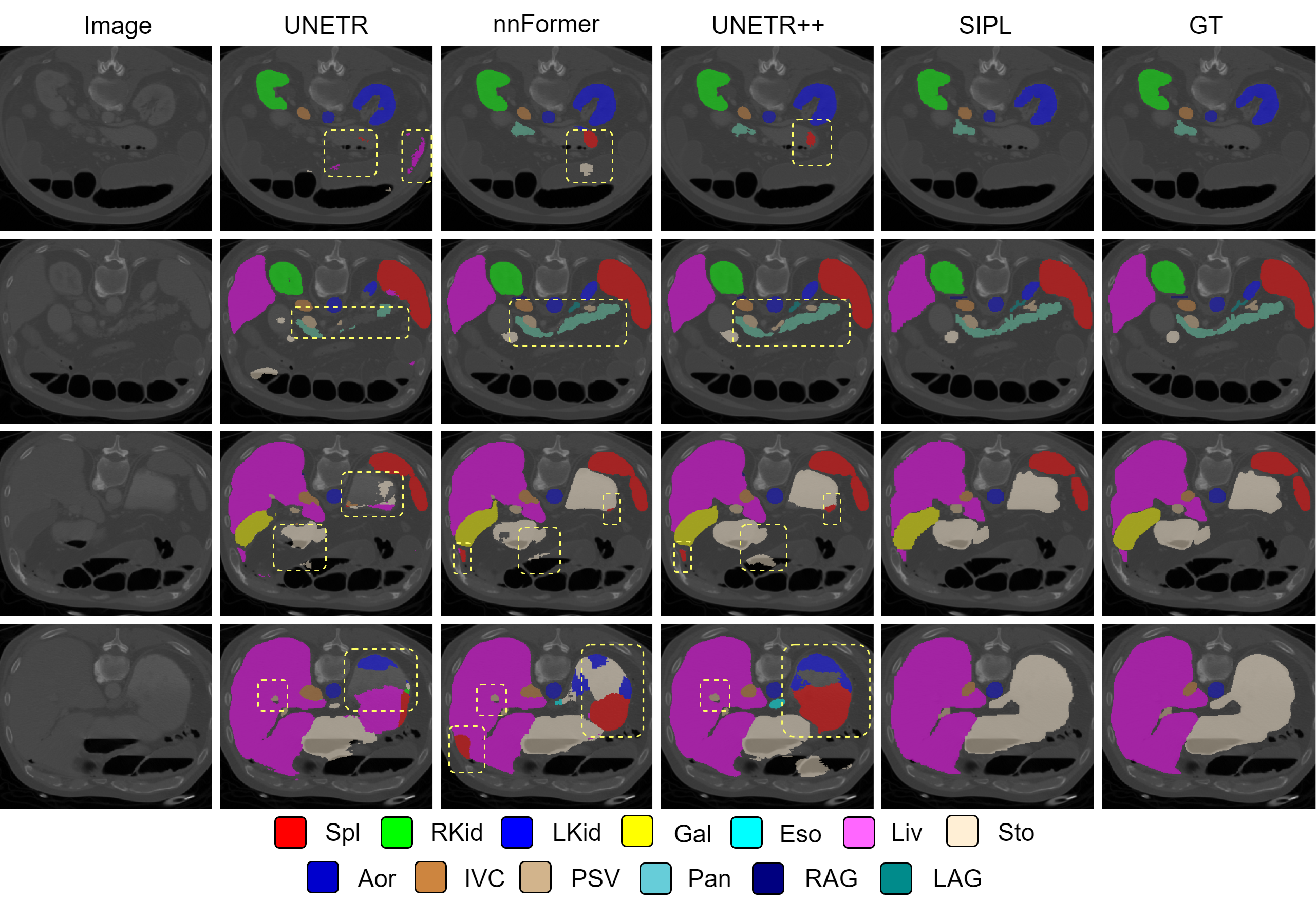}
\caption{Qualitative comparison between SIPL and SOTA models on BTCV dataset. Current methods face challenges in accurately segmenting various organs (highlighted within yellow dashed boxes). In contrast, our proposed framework exhibits commendable segmentation performance.}
\label{fig4:env}
\end{figure*}

\paragraph{Self-supervised Filtering.} 

As mentioned in \cite{yuan2023devil}, generating object queries without proper guidance can introduce a bias towards focusing on the background. To mitigate this issue and prioritize foreground features during training, we introduce a novel self-supervised filtering strategy. This strategy involves comparing the overlapping ratios between the activated regions of each object cluster and the foreground regions of each class. The comparison is performed between the pseudo masks generated by the previous layer $M^{l-1}$ and the pixel-to-cluster assignments at current layer $M^l_{pc}$, which is formulated as,
\begin{equation}
    Overlap(M^{l-1}, M^{l,n}_{pc}) = \mathop{argmax}\limits_{k} \frac{ \sum ( M^{l-1}_k \cap M^{l,n}_{pc})}{\sum M^{l,n}_{pc}},
\end{equation}
where $M^{l,n}_{pc}$ denotes the binarized mask prediction of the $n$-th cluster in $M^l_{pc}$, $ M^{l-1}_k$ refers to the binarized mask prediction of the $k$-th semantic class in $M^{l-1}$. The numerator of $Overlap(,)$ calculates the intersection between the mask of the current cluster and the masks of each semantic class, and $Overlap(,)$ returns the maximum ratio between the cluster-to-class intersection and the area of the related cluster. The underlying assumption is that each cluster represents a specific part of a particular semantic class. Therefore, only clusters with an overlapping ratio higher than the threshold $\tau$ are chosen for query embedding update. In our method, the value of $\tau$ is progressively increased from 0.1 to 0.5 in a linear manner and remains constant after epoch 50, as shown below,
\begin{equation}
    \tau = min(\frac{epoch}{50}, 1.0) \ast 0.4 + 0.1,
\end{equation}
where $epoch$ denotes the index of current training epoch, $min(,)$ returns the smaller value. In order to enhance the quality of the pseudo masks, we introduce an auxiliary loss,
\begin{equation}
    L_{aux}^{l} = 1 - \sum\limits_{k=1}^{K+1} (\sum_i \frac{  Y_{k,i}  M^l_{k,i}}{ Y_{k,i}^2 + {M^l_{k,i}}^2}  + \sum_i Y_{k,i} log M^l_{k,i}  ),
    \label{eq:aux_loss}
\end{equation}
where $Y_{k,i} $, $ M^l_{k,i}$ refer to the results of the $i$-th pixel.
The overall objective for training is formally defined as,

\begin{equation}
L = L_{seg} + \alpha \sum_l L^l_{aux},
\end{equation}
where $\alpha$ is empirically set as 0.05 in our experiments. 

\begin{table*}
    \centering
    \scalebox{0.9}{
    \begin{tabular}{lrrrrrrrrrrrrrr}
        \toprule
        Methods       & Spl & RKid & LKid & Gal & Eso & Liv & Sto & Aor & IVC & PSV & Pan & RAG & LAG & Avg \\
        \midrule
        UNETR         &90.48& 82.51&86.05 &58.23&71.21&94.64&72.06&86.57&76.51&70.37&66.06&66.25&63.04&76.00 \\
        Swin-UNETR    &94.59& 88.97&92.39 &65.37&75.43&95.61&75.57&88.28&81.61&76.30&74.52&68.23&66.02&80.44 \\
        TransBTS      &94.55& 89.20&90.97 &68.38&75.61&96.44&83.52&88.55&82.48&74.21&76.02&67.23&67.03&81.31\\ 
        nnFormer      &94.58& 88.62&93.68 &65.29&76.22&96.17&83.59&89.09&80.80&75.97&77.87&70.20&66.05&81.62 \\
        UNETR++       &\textbf{94.94}& 91.90&93.62 &70.75&77.18&95.95&85.15&89.28&83.14&76.91&77.42&\textbf{72.56}&68.17&83.28 \\
        \midrule
        SIPL (ours)   &94.90&\textbf{94.85}&\textbf{95.23} &\textbf{85.37}&\textbf{81.54}&\textbf{97.14}&\textbf{92.99}&\textbf{89.98}&\textbf{85.77}&\textbf{80.79}&\textbf{83.92}&71.38&\textbf{77.15}&\textbf{87.00}\\  
        \bottomrule
    \end{tabular}}
    \caption{The comparsion results with the state-of-the-art models on BTCV test set. The abbreviations are as follows: Spl: spleen, RKid: right kidney, LKid: left kidney, Gal: gallbladder, Eso: esophagus, Liv: liver, Sto: stomach, Aor: aorta, IVC: the inferior vena cava, PSV: portal and splenic veins, Pan: pancreas, RAG: right adrenal gland, LAG: left adrenal gland.}
    \label{tab1:env}
\end{table*}
\begin{table}
    \centering
    \begin{tabular}{lrr}
        \toprule
        Model             & DSC (\%) & p-values  \\
        \midrule
        UNETR             &73.29    & $< 0.0001$  \\
        nnUNet            &74.31    & $< 0.0001$  \\
        Swin-UNETR         &75.55   & $< 0.0001$  \\
        nnFormer          &77.95    & $< 0.0001$ \\
        UNETR++           &80.68    & $0.0014$  \\
        \midrule
        SIPL (ours)       &\textbf{81.50}   & $--$ \\
        \bottomrule
    \end{tabular}
    \caption{Comparison results with SOTA methods on the Lungs dataset.}
    \label{tab2:env}
\end{table}

\section{Experiments \& Results}
\paragraph{Datasets and Evaluation.}
We conduct experiments on three challenging datasets, including the \textbf{BTCV} dataset~\cite{landman2015miccai}, the \textbf{Lungs} dataset \cite{simpson2019large} and the \textbf{BraTS} dataset \cite{menze2014multimodal}. The \textbf{BTCV} dataset consists of abdominal CT scans from 30 subjects for training and 20 subjects for testing. It includes 13 organs:
\emph{spleen}, \emph{right kidney}, \emph{left
kidney}, \emph{gallbladder}, \emph{liver}, \emph{stomach}, \emph{aorta}, \emph{pancreas}, \emph{esophagus}, \emph{inferior vena cava}, \emph{portal and splenic veins}, \emph{right and left adrenal gland}, respectively. 
The \textbf{Lungs} dataset consists of 63 CT volumes for a binary classification task, specifically focused on delineating lung cancer from the background. The dataset is divided into the training and testing set, with a split ratio of 80:20. 
The \textbf{BraTS} dataset comprises 484 MRI images, each with FLAIR, T1w, T1gd, and T2w channels. The dataset is divided into training, validation, and testing set, with a split ratio of 80:5:15. The target categories for segmentation are whole tumor (WT), enhancing tumor (ET), and tumor core (TC). We utilize the Dice Similarity Coefficient (DSC) as the performance metric for evaluation in all our experiments.

\begin{table}[h]
    \centering
    \begin{tabular}{lrr}
        \toprule
        Model          & DSC (\%)  &p-values \\
        \midrule
        UNETR          &81.2  &0.0003  \\     
        Swin-UNETR     &81.5  & $< 0.0001$\\     
        nnFormer       &82.3  &  0.0025 \\   
        UNETR++        &82.8  & 0.0297 \\     
        \midrule
        SIPL (ours)   &\textbf{83.0} & $--$\\ 
        \bottomrule
    \end{tabular}
    \caption{Comparison results with SOTA methods on the BraTS dataset.}
    \label{tab3:env}
\end{table}

\paragraph{Implementation Details.}
we run all experiments based on Python 3.9, PyTorch 2.1.1 and Ubuntu 22.04. All training procedures have been performed on a single A800 GPU with 80GB memory. 
We report results with $96 \times 96 \times 96$ input size and patch resolution
of $(4, 4, 4)$ for BTCV dataset, where the models are trained for 2k epochs with the default initial learning rate of $4e^{-4}$, momentum of 0.9 and decay of $1e^{-5}$.
For Lungs and BraTS datasets, consistent with \cite{abdelrahman2022unetr++},
the models are trained with input 3D patches of size $128 \times 128 \times 64$ 
for 1k epochs and the learning rate is 0.01 and weight decay is $3e^{-5}$.
The transformer decoder consists of six kMax decoders evenly distributed across feature map with spatial resolutions of $\frac{1}{32}$, $\frac{1}{16}$, and $\frac{1}{8}$. Two kMax decoders are deployed for each resolution. Additionally, the number of object queries is empirically set to 32, and the threshold $\tau$ for query update linearly varies from 0.1 to 0.5 until epoch 50, and remains constant thereafter.

\begin{figure}[h]
\centering
\includegraphics[width=0.49\textwidth]{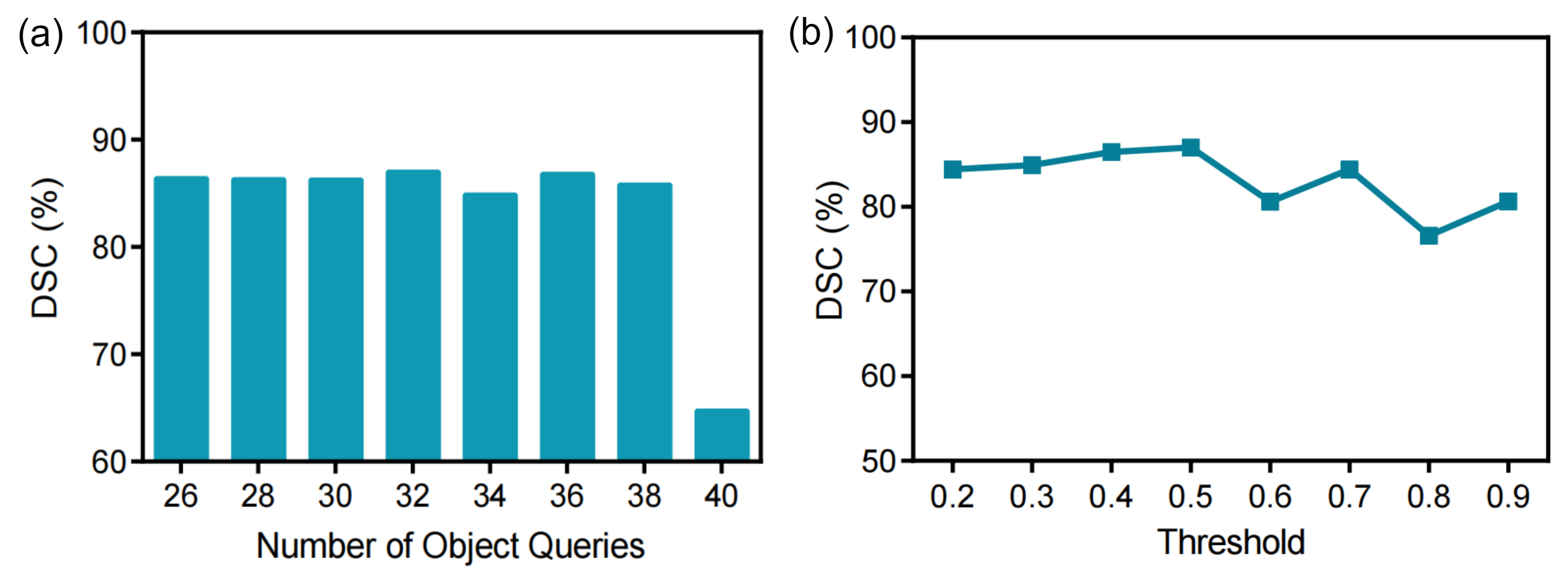}
\caption{SIPL ablation study. (a) The influence of different object query numbers. (b) The influence of different threshold $\tau$.}
\label{fig5:env}
\end{figure}

\begin{table*}[t]
    \centering
    \scalebox{0.9}{
    \begin{tabular}{lrrrrrrrrrrrrrr}
        \toprule
        Model         & Spl & RKid & LKid & Gal & Eso & Liv & Sto & Aor & IVC & PSV & Pan & RAG & LAG & Avg  \\
        \midrule
        - IPL          &90.40&83.99&88.85&66.18&65.39&93.61&88.53&80.27&77.00&58.07&72.36&39.02&47.04&73.13   \\
        - SMG          &93.91&91.80&90.38&80.17&79.63&96.69&91.23&89.96&84.07&77.76&80.17&65.59&72.73&84.16 \\
        \midrule
        SIPL (ours)  &94.90&94.85&95.23&85.37&81.54&97.14&92.99&89.98&85.77&80.79&83.92&71.38&77.15&87.00   \\
        \bottomrule
    \end{tabular}}
    \caption{Illustration on the effectiveness of each component in our method. We remove one component at a time. `- IPL' means removing the IPL module and `- SMG' means removing the SMG module.}
    \label{tab4:env}
\end{table*}
\begin{table*}[t]
    \centering
    \scalebox{0.9}{
    \begin{tabular}{lrrrrrrrrrrrrrr}
        \toprule
        Embedding         & Spl & RKid & LKid & Gal & Eso & Liv & Sto & Aor & IVC & PSV & Pan & RAG & LAG & Avg  \\
        \midrule
        one-hot         &86.16&92.56&90.98&65.38&74.49&95.46&84.41&86.63&76.70&73.58&74.34&47.13&8.26&73.54 \\
        BioBERT         &90.78&92.48&89.95&56.27&63.16&96.58&91.16&88.58&82.46&76.29&80.73&65.48&10.68&75.74 \\
        CLIP            &94.57&94.07&95.05&83.06&77.83&96.99&91.04&90.06&85.75&78.86&83.44&70.86&72.77&85.72 \\
        \midrule
        RanInit (ours)   &94.90&94.85&95.23&85.37&81.54&97.14&92.99&89.98&85.77&80.79&83.92&71.38&77.15&87.00   \\
        \bottomrule
    \end{tabular}}
    \caption{The ablation study on different prototype initialization strategies. Here RanInit represents random initialization.}
    \label{tab5:env}
\end{table*}

\subsection{Comparison with State-of-the-Art Methods}
We compare the segmentation performance between our proposed SIPL framework and the state-of-the-art (SOTA) methods, including nnUNet \cite{isensee2021nnu}, UNETR \cite{hatamizadeh2022unetr}, Swin-UNETR \cite{liu2021swin}, TransBTS \cite{wang2021transbts}, nnFormer \cite{zhou2021nnformer} and UNETR++ \cite{abdelrahman2022unetr++}. 
For fair comparison, all SOTA models were trained/tested on the same dataset splits. The base parameter settings are kept the same as ours.

\paragraph{Multi-organ Segmentation on the BTCV Dataset.}

We provide a detailed comparison with other methods in the testing set of BTCV in Table \ref{tab1:env}.
In comparison, SIPL outperforms the second-best method by $3.72\%$ in terms of mean DSC scores, demonstrating a significant improvement. Specifically, SIPL achieves the highest DSC scores in 11 organs.
Figure \ref{fig4:env} shows a qualitative comparison of our method with existing SOTA approaches on BTCV dataset, where the inaccurate regions are marked with yellow dashed boxes.
The visualization also demonstrates that SIPL obtains favorable segmentation performance than the existing models, achieving the SOTA on the BTCV dataset.
For instance,
In the first row, the \emph{spleen} and \emph{pancreas} are not visible in the slice, but were identified by UNETR, nnFormer, and UNETR++.
In the second row, these models do not perform well in segmenting the pancreas.
In contrast, SIPL accurately segments the \emph{pancreas}  with a superior mean DSC score of $83.92\%$. In addition, UNETR exhibits lower accuracy in segmenting the \emph{right kidney}.
Similarly, in the third row, we observe that UNETR, nnFormer, and UNETR++ struggle to accurately segment the \emph{stomach}, while SIPL achieves the best segmentation performance with a DSC score of $92.99\%$.
Lastly, the listed SOTA methods tend to interfere between adjacent tissues such as the \emph{spleen} when segmenting the \emph{liver} and \emph{stomach} in the fourth row.
These results demonstrate that SIPL outperforms listed SOTA methods with improved boundary delineation in these examples.

\paragraph{Evaluation Results on the Lungs and BraTS Datasets.}

Table \ref{tab2:env} demonstrates the quantitative segmentation performance on the Lungs dataset.
As shown, SIPL outperforms the second best method (UNETR++) by a margin of $0.82\%$ in terms of mean DSC. The p-values from paired t-tests between our method and the listed state-of-the-art (SOTA) methods are all less than 0.05, indicating a statistically significant performance difference between our method and the listed SOTA methods.

Table \ref{tab3:env} displays the comparison results on the BraTS dataset, with the DSC averaged across the three target regions (whole tumor, enhancing tumor, and tumor core). Our method achieves the best segmentation performance, as demonstrated. The p-values (all less than 0.05) further validate that the performance improvements between our method and the listed SOTA methods are statistically significant.



\subsection{Ablation Study}
We now analyze SIPL through a series of ablation studies and all experiments are performed on the BTCV dataset. The DSC scores are leveraged as the default evaluation metric.
\paragraph{Effectiveness of Each Component.}
We evaluate the effectiveness of each component by systematically removing them individually. Table \ref{tab4:env} illustrates the results of these ablation experiments. When the IPL module is absent, our SIPL framework experiences a reduction of $13.87\%$ in overall performance. Similarly, the absence of the SMG module leads to a $2.84\%$ reduction in the average DSC score. These findings highlight the significant impact of the IPL module, which enables the learned prototypes to dynamically adapt to different inputs.

\paragraph{Evaluation on the Influence of Different Prototype Initializations.}
We further show the influence of different initialization strategies of the common prototype proposals (CPPs), including one-hot encoding \cite{zhang2021dodnet}, BioBERT embedding
\cite{yasunaga2022linkbert}, CLIP embedding \cite{radford2021learning} and random initialization embedding (referred as RanInit), as shown in Table \ref{tab5:env}. 
The performance comparison reveals that CLIP embedding, which utilizes image-text pre-training, significantly improves performance compared to traditional one-hot encoding and text-only pre-trained embedding (BioBERT). Surprisingly, random initialization embedding (RanInit) achieves an even larger improvement of $1.28\%$ compared to CLIP embedding. This unexpected result may be attributed to the fact that predefined fixed embeddings could potentially restrict the learning capabilities of the CPPs in our method.



\paragraph{Assessment of the Impact of Object Query Numbers.}

The impact of varying object query numbers is depicted in Figure \ref{fig5:env} (a). It is observed that when the number of object queries is set to 32, the best segmentation performance with a mean DSC of $87.00\%$ is achieved. Results deteriorate when the number of object queries is either too high or too low. A low object query count fails to capture intra-class variation, while a high object query count increases the complexity of pseudo mask generation and may interfere with the main task learning.


\paragraph{Evaluation of the Different Threshold $\tau$.}
We conduct an in-depth analysis of the constant threshold values of $\tau$ in Figure \ref{fig5:env} (b). The results demonstrate that our framework achieves the best performance when the highest threshold is set to $0.5$. Too large or too small threshold values yield inferior outcomes. This is because a large threshold filters out a significant portion of informative pixels during training, hampering the update of query embeddings. Conversely, if the threshold is too small, the learned object embeddings tend to focus on background regions, negatively affecting the segmentation performance.


\section{Conclusion}
In this paper, we propose a novel Self-supervised Instance-adaptive Prototype Learning (SIPL) method for MIS.
Our SIPL framework introduces an Instance-adaptive Prototype Learning (IPL) scheme that adaptively learns prototypes tailored to different inputs.
Moreover, we also design a Self-supervised Mask Generation (SMG) module to hierarchically generate pseudo masks with confidence scores. These pseudo masks have been validated to improve the segmentation performance by generating more discriminative instance-specific prototype proposals.
Experiments and ablation studies validate the effectiveness of each component and demonstrate favorable performance compared to other SOTA methods.
\cite{10183842nnformer}
\cite{wang2022uncertainty}
\cite{10526382unetr}

\section*{Acknowledgements}
This study was supported under the Key Research and Development Program of Zhejiang Province (Grant No. 2023C03192). 
It was also funded by the National Science Foundation of China (Grant No. 62201341).

\section*{Contribution Statement}
Guoyan Liang and Qin Zhou made equal contributions. All
the authors participated in designing research, performing
research, analyzing data, and writing the paper.

\bibliographystyle{named}
\bibliography{ijcai24}

\end{document}


\maketitle

\begin{figure}[t]
\centering
\includegraphics[width=0.48\textwidth]{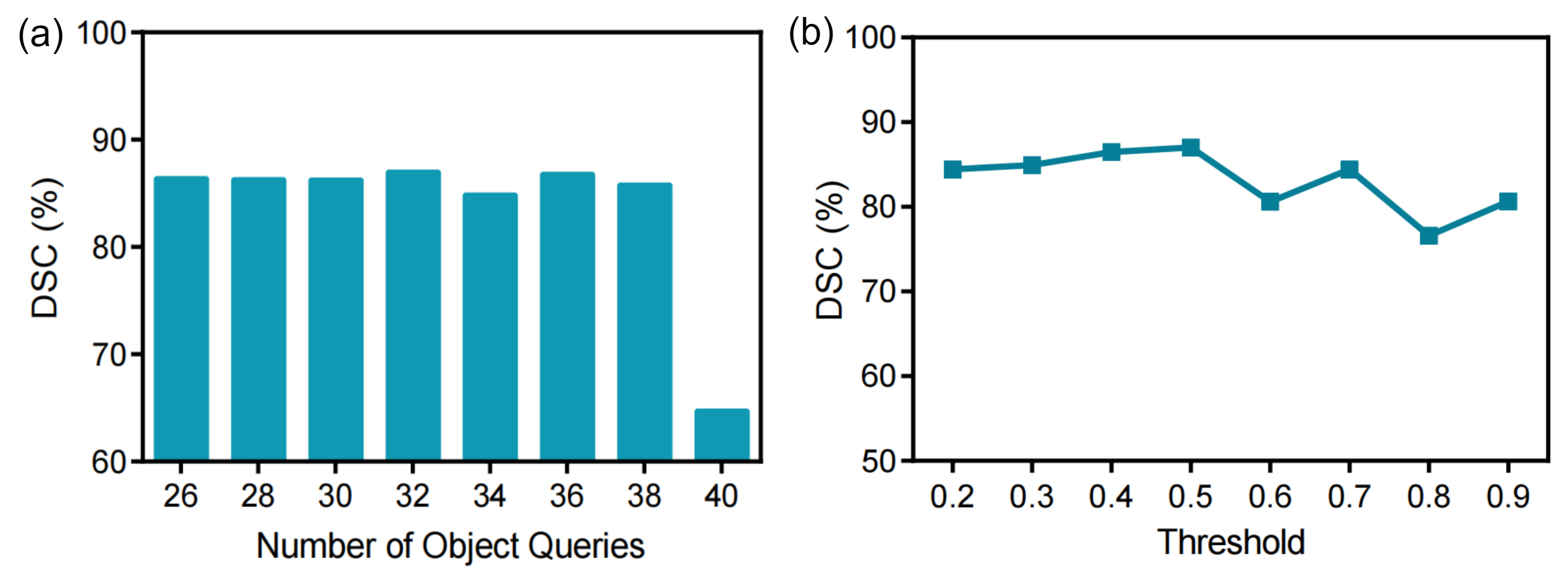}
\caption{Qualitative comparison between SIPL and SOTA model UNETR++ on Lungs dataset.we enhance visibility by enlarging crucial regions (highlighted within yellow dashed boxes) in the images. The inaccurate segmentations are identified with by green dashed boxes. Notably, SIPL outperforms the SOTA models in terms of segmentation accuracy.}
\label{fig5:env}
\end{figure}

\begin{figure}[t]
\centering
\includegraphics[width=0.48\textwidth]{figure/figure7.png}
\caption{Qualitative comparison on BRaTs dataset. Here WT, ET, and TC denote three tumor sub-regions, namely whole tumor, enhancing tumor, and tumor core, respectively.}
\label{fig6:env}
\end{figure}

\paragraph{Evaluation on the Influence of Different Multi-scale Configurations.}
Table \ref{tab7:env} displays our ablation study towards different multi-scale configurations.
Aligning with the previous works \cite{Cheng_2022_CVPR,yu2022k}, using high-resolution features (e.g., a single scale of 1/8) in the Transformer decoder obtains the more benefits.
However, it is proven that will introduce additional computation.
Therefore, we adopt the efficient multi-scale strategy (3 scales) achieves the best performance $87.00\%$ of the mean DSC and does not yield additional computation.
\begin{table*}
    \centering
    \scalebox{0.87}{
    \begin{tabular}{lrrrrrrrrrrrrrr}
        \toprule
        Model         & Spl & RKid & LKid & Gal & Eso & Liv & Sto & Aor & IVC & PSV & Pan & RAG & LAG & Avg  \\
        \midrule
         single scale $\frac{1}{32}$     &93.77&94.74&95.21&84.43&78.97&97.06&92.09&90.32&85.79&77.96&82.93&70.56&74.96&86.06   \\
         single scale $\frac{1}{16}$     &94.28&94.69&94.93&86.16&80.71&96.97&92.42&90.00&85.38&80.33&83.67&71.42&73.25&86.48  \\
         single scale $\frac{1}{8}$      &94.63&94.43&95.29&83.29&79.96&97.14&92.51&90.39&85.54&79.06&82.76&72.57&74.73&86.33  \\
        \midrule
        SIPL (3 scales)   &94.90&94.85&95.23&85.37&81.54&97.14&92.99&89.98&85.77&80.79&83.92&71.38&77.15&87.00  \\
        \bottomrule
    \end{tabular}}
    \caption{Ablation study on the influence of different multi-scale configurations.}
    \label{tab7:env}
\end{table*}

\begin{table*}
    \centering
    \scalebox{0.9}{
    \begin{tabular}{lrrrrrrrrrrrrrr}
        \toprule
        Cluster Number        & Spl & RKid & LKid & Gal & Eso & Liv & Sto & Aor & IVC & PSV & Pan & RAG & LAG & Avg  \\
        \midrule
        26         &93.10&94.69&95.17&84.29&79.23&96.96&92.41&90.47&85.62&79.77&83.32&71.60&76.72&86.41 \\
        28         &93.89&94.86&95.28&83.45&80.38&97.11&92.63&90.24&86.24&79.61&83.10&71.28&74.19&86.33  \\
        30         &94.83&94.66&95.30&82.74&79.99&97.12&92.64&90.51&86.02&79.38&83.34&70.49&74.77&86.29  \\
        32 (ours)   &94.90&94.85&95.23&85.37&81.54&97.14&92.99&89.98&85.77&80.79&83.92&71.38&77.15&87.00  \\
        34         &93.29&94.14&94.87&83.53&71.35&96.90&92.41&89.85&83.22&79.59&83.10&70.93&70.21&84.88  \\
        36         &94.98&95.13&95.26&83.89&80.92&97.17&93.18&90.51&86.11&80.60&84.49&70.10&76.31&86.82  \\
        38         &94.37&93.82&95.22&82.36&78.42&97.05&91.69&89.34&85.13&79.47&83.16&69.60&76.01&85.82  \\
        40         &85.11&84.82&87.85&49.76&66.16&93.41&82.48&83.69&74.34&66.76&67.20&0.00&0.00&64.74 \\
        \bottomrule
    \end{tabular}}
    \caption{Ablation study of the number of different soft cluster.}
    \label{tab6:env}
\end{table*}

\begin{table*}
    \centering
    \scalebox{0.9}{
    \begin{tabular}{lrrrrrrrrrrrrrr}
        \toprule
        Threshold         & Spl & RKid & LKid & Gal & Eso & Liv & Sto & Aor & IVC & PSV & Pan & RAG & LAG & Avg  \\
        \midrule
         0.3          &90.90&92.52&93.36&82.56&78.65&96.06&90.86&89.47&84.10&79.86&81.34&70.11&74.48&84.94  \\
         0.4          &94.72&94.65&95.27&83.57&79.58&97.05&92.55&90.43&85.85&79.35&83.53&71.39&76.03&86.46  \\
         0.5 (ours)   &94.90&94.85&95.23&85.37&81.54&97.14&92.99&89.98&85.77&80.79&83.92&71.38&77.15&87.00  \\
         0.6          &91.09&91.98&93.99&75.71&69.08&95.94&88.36&87.36&80.93&76.21&76.80&63.77&56.40&80.59  \\
         0.7          &94.46&87.57&94.67&85.24&74.95&96.81&91.51&89.47&84.54&80.00&83.41&60.97&73.95&84.43  \\
         0.8          &86.80&84.23&83.56&79.48&71.86&95.28&87.29&84.86&76.35&72.32&69.36&55.88&48.49&76.60  \\
        \bottomrule
    \end{tabular}}
    \caption{Ablation on the influence of different threshold $\tau$.}
    \label{tab8:env}
\end{table*}